# Maximizing solubility in rock salt high-entropy oxides

Matthew Furst[1], Joseph Petruska[1], Dhiya Srikanth[1], Jacob T. Sivak[2], Susan B. Sinnott[1,2,3], Christina M. Rost[4], Jon-Paul Maria[1], and *Saeed S. I. Almishal[1]

[1]Department of Materials Science and Engineering, The Pennsylvania State University, University Park, PA 16802, USA
[2]Department of Chemistry, The Pennsylvania State University, University Park, PA 16802, USA
[4]Institute for Computational and Data Sciences, The Pennsylvania State University, University Park, PA 16802, USA
[4]Department of Materials Science and Engineering, Virginia Polytechnic Institute and State University, Blacksburg, VA 24061, USA

**Corresponding Authors:** Saeed S. I. Almishal saeedsialmishal@gmail.com

## Abstract

To explore and quantitatively map the cation-size mismatch solubility limits in high-entropy oxides (HEOs), we report on $Ca^{2+}$ substitution in prototypical MgCoNiCuZnO, because while isovalent, $Ca^{2+}$ is 38% larger than its partners' average ionic radii. Using the thermodynamics-grounded bond-length distribution descriptor, we identify $Ca^{2+}$–$Cu^{2+}$ interactions as the primary prospective lattice destabilizer. Bulk synthesis confirms only 4% Ca solubility with Cu at 950°C, modestly rising to 5% after Cu removal at 1150°C. We then employ far-from-equilibrium pulsed-laser deposition to investigate metastable solubility; epitaxial films incorporate 10% Ca with Cu and a full 20% Ca without, doubling and quadrupling the respective bulk limits. This Ca uptake additionally enables deterministic lattice-parameter control via Ca concentration. Overall, our results demonstrate both the extended solubility that is possible in HEO systems, particularly when accessing metastable states through quenching from high-energy plasma, and that the specific constellation of solid solvent cations can be rationally engineered to minimize bond-length distributions when largely misfit cations are added, thus expanding the accessible compositional space.

## Introduction

High-entropy oxides (HEOs), as well as their entropy-stabilized subset, leverage configurational entropy – which dominates and enhances the thermal energy of mixing ($T\Delta s_{mix}$) – to overcome a positive enthalpy of mixing ($\Delta h_{mix}$). This lowers the chemical potential for stabilizing a multi-component solid solution ($\Delta \mu = \Delta h_{mix} - T\Delta s_{mix}$), particularly at elevated temperatures. A hallmark example is the 2015 HEO $Mg_{1/5}Co_{1/5}Ni_{1/5}Cu_{1/5}Zn_{1/5}O$ (MgCoNiCuZnO for brevity), which stabilizes a single-phase rock salt structure from otherwise immiscible and non-isostructural oxides[1–6]. Configurational entropy alone, however, cannot guarantee phase stability; a positive $\Delta h_{mix}$ can pose a substantial barrier to single-phase stability when cations differ significantly in size or valence[3,5,7–9]. Classical, enthalpy-based solubility criteria, inspired by Hume-Rothery[10] and Pauling rules[11], therefore remain essential to consider within HEOs. Achieving a stable single-phase in practice still demands (i) minimizing lattice strain via close cation-size matching and (ii) preserving charge neutrality through compatible valence states. Figure 1(a) illustrates the cation size matching using ionic radii: the divalent cations in MgCoNiCuZnO span a narrow window, with the largest deviation ($Ni^{2+}$ vs $Co^{2+}$) being only ~8%.

A periodic table survey indicates few cation sets which satisfy the size and valence constraints needed to neatly demonstrate entropic stability as in MgCoNiCuZnO; expanding HEO compositional space therefore requires pushing beyond these more comfortable regimes. Divalent $Mn^{2+}$ and $Fe^{2+}$ and trivalent $Sc^{3+}$ and $Cr^{3+}$ can be accommodated by controlling oxygen chemical potential[9] and by defect compensation[12,13] respectively; these sit within a relatively narrow ionic radii dispersion. To expand the single-phase formulation space, we now consider whether the lattice can endure far more extreme size misfit cations. $Ca^{2+}$ is the ideal probe to test this hypothesis: it retains the 2+ charge and rock salt preference of MgCoNiCuZnO yet carries a radius roughly 38% larger than the host cation average radius (Figure 1(a)). Testing solubility limits with $Ca^{2+}$ therefore allows us to map the limits for size misfit cations, quantify the accompanying energetic penalties, and identify pathways to overcome these limitations to maximize misfit solubility (hereafter, we omit the 2+ charge, though it remains implied throughout). Ionic radii mismatch alone is an incomplete perspective, however, as additional local lattice strain can be induced by electronically-driven distortions like the CuO Jahn-Teller elongation[14–16]. To capture this, we shift our focus from simple ionic radii to the relaxed first near-neighbor cation-anion bond length standard deviation ($\sigma_{bonds}$) for ternary rock salt oxides. Larger values indicate cations that introduce more significant local relaxations and distortions in the lattice (details reported in Ref [17]). Figure 1(b) summarizes such ternary $\sigma_{bonds}$ retrieved from Ref.[17] to highlight that while Mg, Co, Ni, and Zn combinations all have $\sigma_{bonds} < 0.03$ Å, systems containing Ca or Cu possess considerably larger values. The Ca–Cu pair indeed exhibits the highest $\sigma_{bonds}$, which we attribute to the combined impact of Ca's large size and Cu's Jahn-Teller distortion.

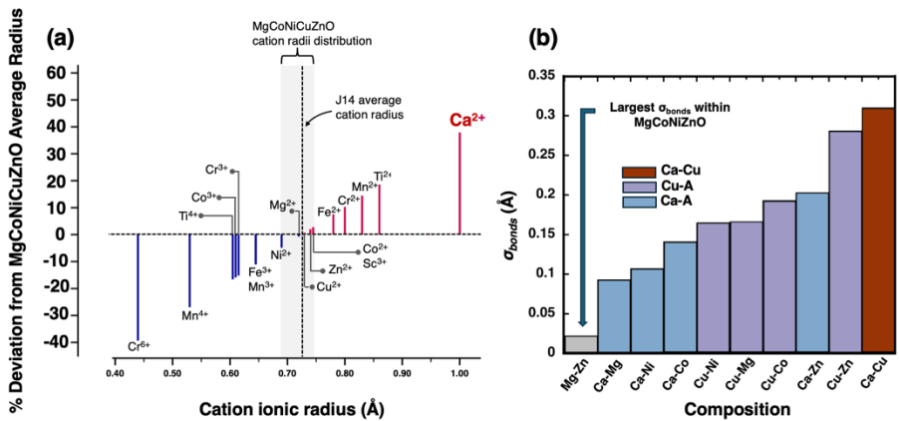

**Figure 1: Cation size and bond-length distribution descriptors:** (a) Percent deviation from MgCoNiCuZnO average cation radius versus cations ionic radii, (b) Standard deviation of first near-neighbor bond lengths ($\sigma_{bonds}$) bar chart of $A_{1/2}A'_{1/2}O$ rock salt ternary oxides from the cation cohort: {Ca, Mg, Co, Ni, Cu, Zn}.

It follows that Ca's oversized ionic radius prevents its equimolar dissolution in MgCoNiCuZnO using bulk synthesis. We therefore examine whether compositional tuning (lower Ca concentrations and Cu removal) and synthesis energetics (high-temperature and far-from-equilibrium routes) can overcome this stability barrier and expand Ca's solubility window. Specifically, we propose that a narrow Ca range exists where the combined mixing enthalpy and elastic strain can be minimized to stabilize single-phase $Ca_x(MgCoNiCuZn)_{1-x}O$. Removing Cu, whose Jahn-Teller distortions amplify the Ca-induced penalty, should shift this window toward higher Ca concentrations in $Ca_x(MgCoNiZn)_{1-x}O$. Although the four-component host carries less configurational entropy, the absence of Cu lifts the 950 °C reduction/melting ceiling, so firings up to ~1150 °C can reclaim the original $T\Delta s_{mix}$ and drive additional solubility. To exceed even these bulk limits, we grow thin films by pulsed-laser deposition (PLD): the high-temperature plume and rapid quench at low substrate temperatures should kinetically arrest Ca-rich rock salt metastable phases unreachable under equilibrium[2,12,15,18–20]. Together, these equilibrium and far-from-equilibrium routes pinpoint the true size-misfit tolerance of rock-salt HEOs and establish practical levers – composition, temperature, and quench – to unlock large-radius cation incorporation.

## Main

### 1    Ca solubility in MgCoNiCuZnO and MgCoNiZnO: Equilibrium synthesis

Incorporating Ca into the prototypical MgCoNiCuZnO in bulk is the initial experimental priority. We have previously explored Ca concentrations up to 16.6% (detailed in Supporting Information Notes 1 and 2), finding that compositions even as low as 6.2% Ca yield a two phase-mixture after reaction for 18 hours at 950ºC[17]. Here, we focus on a narrower Ca range to examine the transition from two-phase to single-phase solid solution and its associated change in lattice parameter. Figure 2(a) presents Bragg-Brentano High-Definition X-Ray Diffraction (BBHD XRD) scans for reacted $Ca_x(MgCoNiCuZn)_{1-x}O$ at 950ºC, with Ca concentration in the 0% to 5.5% range.

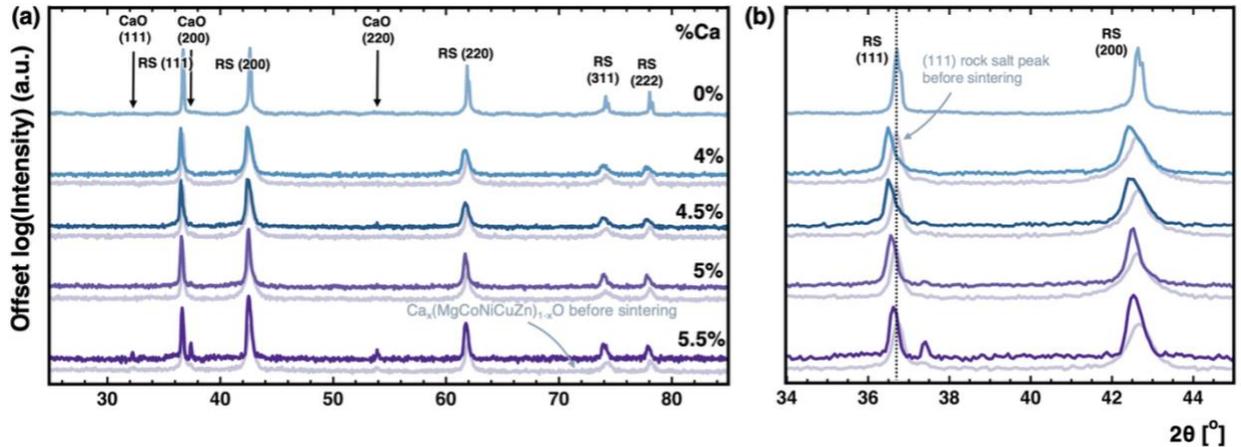

**Figure 2: Cu-containing compositions bulk synthesis:** (a) Bragg-Brentano High-Definition X-ray diffraction patterns with %Ca solubility in MgCoNiCuZnO ranging from 5.5% to 4%, (b) an enlarged scan of the (111) and (002) peaks, revealing the systematic lattice expansion produced by Ca incorporation into the high-entropy rock-salt lattice. For each composition, the diffraction trace of the corresponding unsintered precursor powder, xCaO+(1-x)(MgCoNiCuZn)O, is shown directly beneath the sintered pattern. Arrows mark the expected CaO peaks; the CaO (200) peak, for example, is prominent in the reacted 5.5 at % Ca sample. Note these peaks are barely discernible in the precursor powders due to their lower crystallinity prior to sintering. Any apparent peak splitting is attributed to Cu $K_\alpha$ and $K_\beta$ radiation from the X-ray source.

Gray faded scans represent pre-reacted MgCoNiCuZnO mixed with respective CaO concentrations prior to a high-temperature reaction for comparison. At Ca concentrations ≥5.5%, we observe two clear peak sets indicating rock salt MgCoNiCuZnO, and CaO. As the Ca content decreases to 5% and 4.5%, the intensity of the CaO peaks diminishes progressively until 4% Ca where diffraction suggest a single phase but with larger lattice parameter values. Such a shift is expected due to the larger Ca ion expanding the average lattice parameter of the high-entropy matrix even at modest concentrations. To illustrate this trend, Figure 2(b) magnifies the rock-salt (111) and (200) reflections. Notably, the 4% Ca sample shows the largest lattice parameter in the series, signifying the highest Ca solubility. As the Ca content rises from 4% to 5.5%, the lattice parameter contracts and a secondary CaO-rich phase re-emerges, indicating that these higher concentrations exceed the solubility limit at our synthesis conditions.

We next remove Cu to eliminate both its associated Jahn-Teller distortions and the 950 °C firing limit. Figure 3(a) presents BBHD XRD patterns for Cu-free $Ca_x(MgCoNiZn)_{1–x}O$ compositions reacted at 1150 °C, with additional data for Ca concentrations up to 20% provided in Supporting Information Figure S3. The narrower peaks and clear splitting of the (220), (311), and (222) reflections indicate enhanced crystallinity, which we attribute to both the removal of Cu-induced Jahn–Teller distortions and the elevated reaction temperature. Down to 5.5% Ca, a weak CaO (200) reflection is still present alongside the rock salt peaks. In contrast, at Ca concentrations ≤5%, we observe a single set of rock salt peaks, consistent with a single-phase HEO. Magnifying the rock-salt (200) peaks in Figure 4(b), we see that the lattice parameter increases steadily from 0% to 5% Ca but plateaus beyond that, further supporting saturation and the onset of a CaO-rich secondary phase. Notably, this solubility threshold is just slightly higher than in the Cu-containing system (increasing from 4% to 5%), emphasizing that size-misfit cations remain challenging to incorporate at relatively larger amounts under equilibrium synthesis.

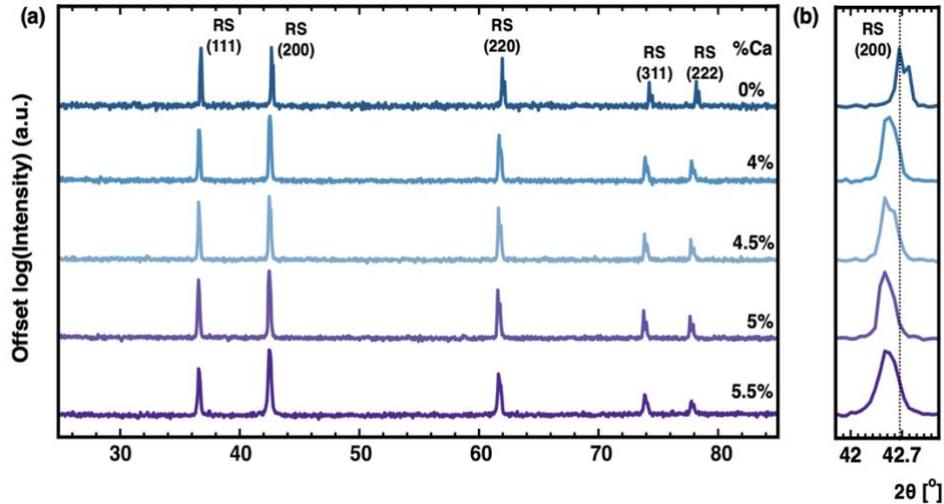

**Figure 3: Cu-free compositions bulk synthesis:** (a) Bragg–Brentano High-definition X-ray diffraction patterns for MgCoNiZnO with Ca solubility ranging from 0% to 5.5%. (b) Enlarged view of the (002) reflections, highlighting the lattice parameter shift induced by Ca incorporation after 18 hours of sintering at 1150°C. Any apparent peak splitting is attributed to Cu $K_\alpha$ and $K_\beta$ radiation from the X-ray source.

## 2 Ca solubility in MgCoNiCuZnO and MgCoNiZnO: Far-from-equilibrium synthesis

To push beyond equilibrium solubility limits, we employ far-from-equilibrium kinetic pathways enabled by PLD, where high-energy ablation conditions should promote higher Ca concentrations into rock salt HEO epitaxial films. We grow the films at 1.87 J/cm$^2$ laser fluence and 400°C substrate temperature, well below bulk synthesis temperatures to enable rapid quenching from the plasma plume and metastabilize the high-symmetry, high-entropy phase[2,15]. Supporting Information Note 4 details identifying and establishing favorable growth conditions.

We begin with Cu-containing Ca$_x$(MgCoNiCuZn)$_{1-x}$O films, systematically increasing the Ca concentration. Figure 4(a) shows wide 2θ BBHD XRD scans for each composition. Across all films, only (002) and (004) reflections are observed, indicating epitaxial growth on MgO (001). As in bulk ceramics, we expect the out-of-plane lattice parameter to increase with higher Ca content. Indeed, the lattice parameter increases up to 10% Ca, beyond which it decreases indicating we are approaching the solubility limit. While still constrained, this limit exceeds the Ca incorporation achievable under equilibrium synthesis by more than twofold. In Figure 4(a), we also observe a broad, low-intensity peak to the left of the main film peak which we attribute to Ca-rich secondary phases with large lattice distortions. To better resolve the (002) film peaks, Figure 4(b) shows the corresponding high-resolution XRD scans. From 0-10% Ca, Pendellösung fringes attest to high crystalline fidelity and smooth interfaces. As Ca content increases, these fringes progressively diminish, consistent with increasing lattice distortions from the size-misfit cation. Between 5% and 10% Ca, additional weak reflections appear that do not correspond to known orientations and may represent satellite peaks. At 13.5% and 16.6% Ca, the high-resolution scans reveal two overlapping peaks not resolved in the wide-range BBHD scans. One is the relatively narrow main $Ca_x(MgCoNiCuZn)_{1-x}O$ peak, though with reduced crystalline quality. The other is a broad shoulder on the low-angle side, again suggesting the presence of Ca-rich secondary phases with large lattice distortions.

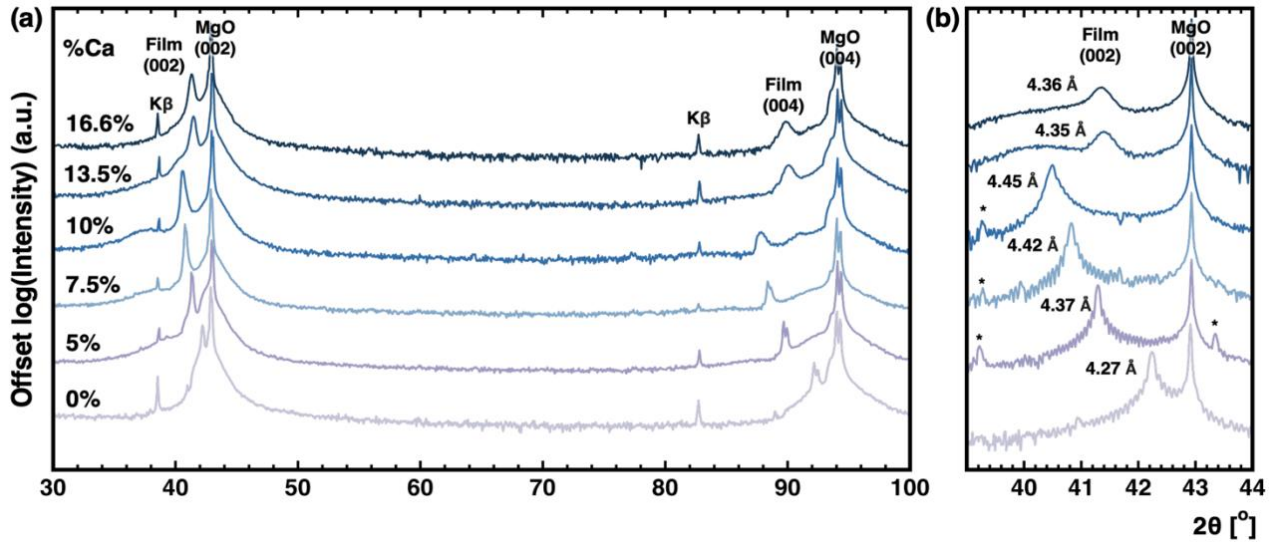

**Figure 4: Cu-containing compositions thin film synthesis**: (a) Bragg-Brentano High-Definition X-ray diffraction scans highlighting CaO solubility in MgCoNiCuZnO ranging from 0-16.6%, (b) High-resolution scans using a hybrid monochromator on the incident beam and crystal analyzer detector, focusing on the (002) reflections of both film and substrate for Ca incorporation levels from 0% to 16%. Film lattice parameters, extracted from peak positions, are labeled in angstroms adjacent to their corresponding reflections. Apparent peak splitting is attributed in BBHD to Cu $K_\alpha$ and $K_\beta$ radiation from the X-ray source.

To untangle Cu's impact on Ca far-from-equilibrium solubility, we remove Cu from the composition, analogous to Figure 3, and grow Cu-free $Ca_x(MgCoNiZn)_{1-x}O$ films under the same conditions used in Figure 4. We show the BBHD XRD patterns for this series in Figure 5(a). As in the Cu-containing series (Figure 4a), only the (002) and (004) film reflections appear, confirming epitaxial growth on MgO. In contrast to the Cu case, the out-of-plane lattice parameter increases smoothly up to the equimolar composition, demonstrating a markedly higher Ca solubility. High-resolution XRD scans of the (002) peak in Figure 5(b) reveal more nuanced features: at 5-7.5% Ca, two superimposed peaks emerge. One is narrower and shifts to lower 2θ values relative to the 0% film, while the other is broader and lower in intensity, consistent with a distorted Ca-rich secondary phase. At Ca concentrations ≥10%, these features consolidate into a single peak, indicating recovery of a single-phase film. However, Pendellösung fringes fade with increasing Ca content, pointing to reduced crystalline quality caused by lattice distortions from the size-misfit Ca cations.

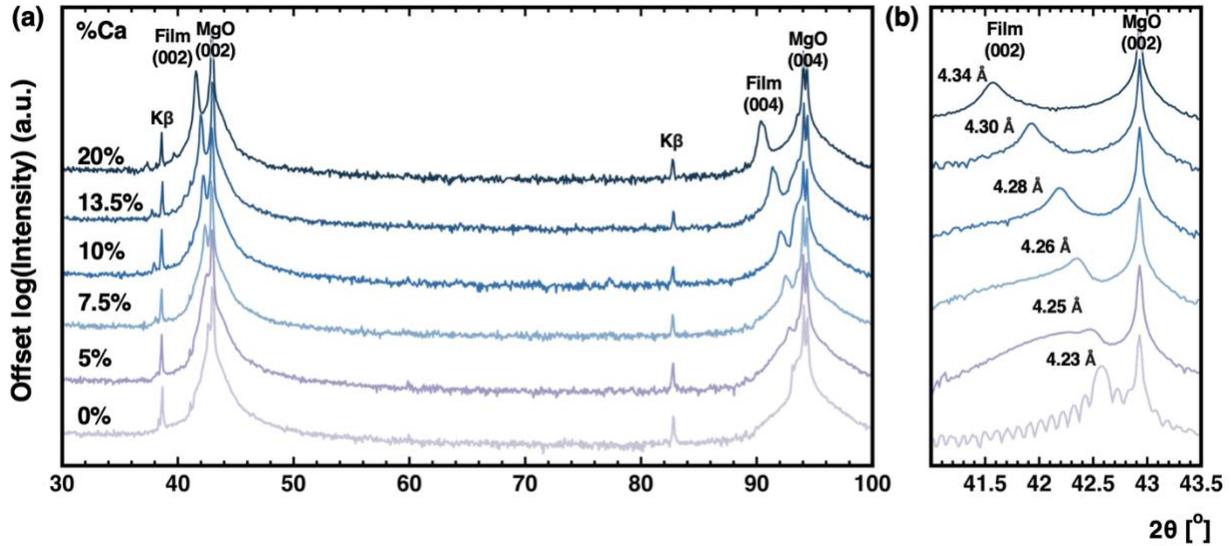

**Figure 5: Cu-free compositions thin film synthesis**: (a) Bragg–Brentano High-Definition X-ray diffraction scans showing Ca incorporation in MgCoNiZnO with Ca content ranging from 0% to 20%, (b) high-resolution X-ray diffraction scans using a hybrid monochromator and crystal analyzer detector, focused on the (002) peaks of both film and substrate. Film lattice parameters, extracted from peak positions, are labeled in angstroms adjacent to their corresponding peaks. Apparent peak splitting is attributed in BBHD to Cu $K_\alpha$ and $K_\beta$ radiation from the X-ray source. Small, narrow peaks appearing at lower 2θ angles to the left of the substrate $K_\beta$ peak originate from the film $K_\beta$ signal.

## Discussion and Outlook

Our combined equilibrium and far-from-equilibrium experiments converge on four guiding conclusions: (1) Ca solubility is capped at ~4% in bulk ceramics and ~10% in PLD films for the Cu-containing MgCoNiCuZnO system; Ca-rich secondary phases appear and Pendellösung fringes fade beyond these thresholds, (2) removing Cu alleviates Jahn–Teller distortions and broadens the solubility window; modestly in bulk (to ~5%) and dramatically in films, where single-phase growth persists up to the equimolar composition, (3) far-from-equilibrium PLD pushes Ca concentrations far beyond bulk limits but simultaneously amplifies the trade-off between kinetic trapping and strain energy, and (4) irrespective of Cu presence, larger Ca content increases the lattice parameter and lowers the crystalline quality. Conclusion (2) underscores a striking message: despite having the smallest deviation from the average ionic radius in MgCoNiCuZnO (Figure 1(a)), Cu introduces the most severe lattice strain with Ca, as quantified by the $\sigma_{bonds}$ metric (Figure 1(b)). This interplay come into sharp focus when comparing two otherwise identical 7.5% Ca films: one Cu-containing, one Cu-free (Figure 6(a)), neither of which can be stabilized as single-phase materials under equilibrium conditions. The Cu-containing film exhibits a 3.5% larger out-of-plane lattice parameter than its Cu-free counterpart, yet retains pronounced Pendellösung fringes, indicating higher crystalline fidelity. We attribute this increase in lattice parameter to lattice distortions introduced by Cu's Jahn-Teller effects combined with Ca's size in agreement with Figure 1(b). We hypothesize that the enhanced fringe fidelity likely arises from the ~13% higher configurational entropy in $Ca_{0.075}(MgCoNiCuZn)_{0.925}O$ compared to $Ca_{0.075}(MgCoNiZn)_{0.925}O$ (based on ideal entropy of mixing), which mitigates the added strain energy and assists in stabilizing the rock-salt phase.

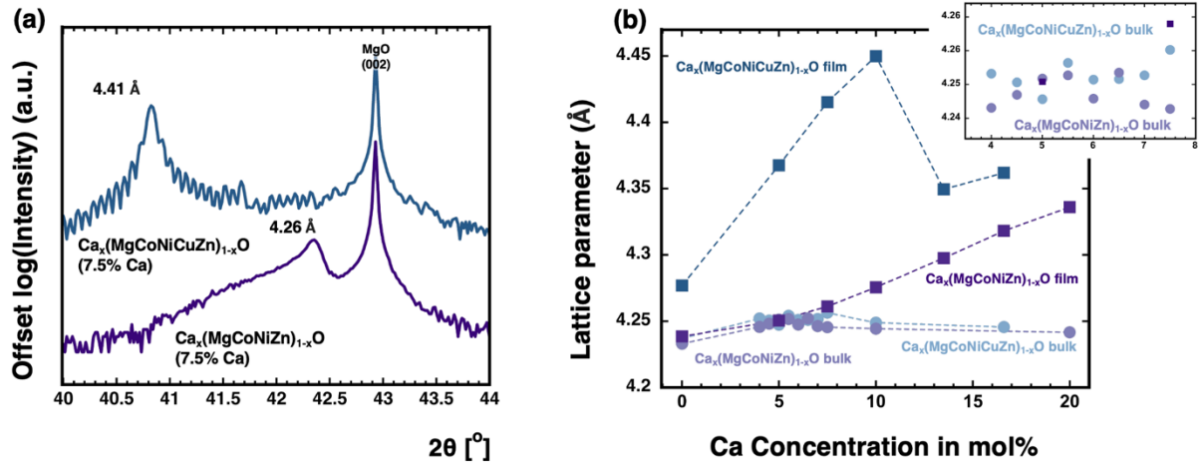

**Figure 6: Lattice parameter trends with Cu inclusion and Ca concertation:** (a) high-resolution X-ray diffraction scans using a hybrid monochromator and crystal analyzer detector, focused on the (002) film peaks for 7.5% Ca in MgCoNiCuZnO (top) and MgCoNiZnO (bottom). Film lattice parameters are added in angstroms adjacent to their respective peaks. (b) Evolution of the out-of-plane lattice parameter with Ca concentration for both bulk and thin-film compositions of $Ca_x(MgCoNiCuZn)_{1-x}O$ and $Ca_x(MgCoNiZn)_{1-x}O$, with zoomed-in inset for the bulk values. Lines are added to guide the eye and clarify composition-dependent trends.

Figure 6(b) distills the lattice parameter evolution across all compositions. In bulk ceramics, the lattice expands modestly with increasing Ca, closely tracking Vegard's law, and plateauing with no clear trend past their respective Ca solubility limits (~4-5%). In contrast, thin films exhibit substantially larger out-of-plane lattice parameters that rise sharply with Ca content. Cu-containing films expand by 4.2% over the first 10% Ca before saturating – clearly reflecting the Cu–Ca interplay described above. Cu-free films, by comparison, expand almost linearly, by 2.6% over 20% Ca range, showing that once Jahn–Teller activity is absent, Ca affords precise and predictable control over the out-of-plane lattice parameter.

Broadly, we reveal that cation size mismatch, exemplified by Ca, presents a more formidable solubility barrier in high-entropy oxides than previously appreciated, surpassing challenges posed by isostructural or aliovalent cations such as $Zn^{2+}$ and $Sc^{3+}$[12]. Nonetheless, we demonstrate that sufficiently high growth energetics, coupled with rapid quenching from the plasma at low substrate temperatures, can enable the misfit cation to be incorporated in equimolar proportions. The resulting solid solutions exhibit large lattice strains, high defect densities, and pronounced chemical disorder while retaining exceptional crystallinity. These findings position size-mismatched cations as powerful tools to tune carrier dynamics, background conductivity, and defect excitation landscapes in entropy-stabilized materials, paving the way for memristive devices driven by electronic rather than ionic mechanisms[21,22]. Moreover, our synthesis approach can be extended to incorporate size-mismatched cations with targeted functionalities, such as magnetism or optical fluorescence, opening the door to a broader compositional space and expanded opportunities for functional high-entropy materials[2,18,23–28].

## Methods

*Bulk Synthesis*

Depending on composition, we combine MgO (Sigma-Aldrich, 342793), CoO (Sigma-Aldrich, 343153), NiO (Sigma-Aldrich, 203882), CuO (Alfa Aesar, 44663), ZnO (Sigma-Aldrich, 96479), and CaO (Sigma-Aldrich, 208159). We mix and mill the powders with 5 mm yttrium-stabilized zirconia media for 2.5 hours. For bulk studies, we press the powder into 1.25 cm diameter pellets at 60-100 MPa (Carver Laboratory Press) for 30 seconds. To make bulk targets for pulsed laser deposition, we press the powder into 2.5 cm diameter pellets at 60-100 MPa for 60 seconds. We fire a typical $Ca_x(MgCoNiCuZn)_{1-x}O$ sample at 950-1000°C, and a typical $Ca_x(MgCoNiZn)_{1-x}$ sample at 1150°C. All bulk samples are fired in a box furnace in ambient air conditions. We quench all samples from ~700 °C in air to prevent low-temperature phase segregation. We verify structural composition via X-ray diffraction (Panalytical Empyrean) using $2\theta-\theta$ Bragg-Brentano HD scans with a PIXcel3D detector and identify phases with PANalytical HighScore. We monitor cation compositions before and after sintering by X-ray fluorescence (Panalytical Epsilon 1).

*Thin-film synthesis*

We anneal (100)-MgO substrates on the chamber heater at 800°C for 20 minutes, followed by 20 minutes at the growth temperature, typically 400°C (Supporting Information Note 4). We grow under 50 sccm flow of $O_2$ gas maintaining an internal pressure of 50 mTorr. We set the laser to 2000 pulses at a frequency of 10 Hz, maintaining a typical fluence of 1.87 J/cm$^2$. After growth, all films are rapidly removed from the chamber to prevent low-temperature phase segregation. We verify structural composition via X-ray diffraction (Panalytical Empyrean) using $2\theta-\theta$ Bragg-Brentano HD scans with a PIXcel3D detector. We perform high-resolution scans with a 2xGe hybrid monochromator on the incident side and a another 2xGe crystal analyzer with a proportional detector on the diffracted side.

## Acknowledgments

The authors gratefully acknowledge support from NSF MRSEC DMR-2011839. The authors also acknowledge Kaylin Lamaute for valuable assistance with bulk-ceramic processing.

## References

1. Rost, C. M. *et al.* Entropy-stabilized oxides. *Nat Commun* **6**, 8485 (2015).
2. Almishal, S. S. I. *et al.* Untangling individual cation roles in rock salt high-entropy oxides. *Acta Materialia* **279**, 120289 (2024).
3. Kotsonis, G. N. *et al.* High-entropy oxides: Harnessing crystalline disorder for emergent functionality. *Journal of the American Ceramic Society* **106**, 5587–5611 (2023).
4. Rost, C. M. Entropy-Stabilized Oxides: Explorations of a Novel Class of Multicomponent Materials. (2016).

5. McCormack, S. J. & Navrotsky, A. Thermodynamics of high entropy oxides. *Acta Materialia* **202**, 1–21 (2021).
6. Sarkar, A., Breitung, B. & Hahn, H. High entropy oxides: The role of entropy, enthalpy and synergy. *Scripta Materialia* **187**, 43–48 (2020).
7. Pitike, K. C., KC, S., Eisenbach, M., Bridges, C. A. & Cooper, V. R. Predicting the Phase Stability of Multicomponent High-Entropy Compounds. *Chem. Mater.* **32**, 7507–7515 (2020).
8. Pu, Y. *et al.* (Mg,Mn,Fe,Co,Ni)O: A rocksalt high-entropy oxide containing divalent Mn and Fe. *Sci. Adv.* **9**, eadi8809 (2023).
9. Almishal, S. S. I. *et al.* Thermodynamics-Inspired High-Entropy Oxide Synthesis. Preprint at https://doi.org/10.48550/arXiv.2503.07865 (2025).
10. Hume-Rothery, W. & Powell, H. M. On the theory of super-lattice structures in alloys. *Zeitschrift für Kristallographie-Crystalline Materials* **91**, 23–47 (1935).
11. Pauling, L. The sizes of ions and the structure of ionic crystals. *Journal of the American Chemical Society* **49**, 765–790 (1927).
12. Kotsonis, G. N., Rost, C. M., Harris, D. T. & Maria, J.-P. Epitaxial entropy-stabilized oxides: growth of chemically diverse phases via kinetic bombardment. *MRS Communications* **8**, 1371–1377 (2018).
13. Niculescu, G. E. *et al.* Local structure maturation in high entropy oxide (Mg,Co,Ni,Cu,Zn)1-(Cr,Mn)O thin films. *Journal of the American Ceramic Society* **108**, e20171 (2025).
14. Berardan, D., Meena, A. K., Franger, S., Herrero, C. & Dragoe, N. Controlled Jahn-Teller distortion in (MgCoNiCuZn)O-based high entropy oxides. *Journal of Alloys and Compounds* **704**, 693–700 (2017).
15. Almishal, S. S. I. *et al.* Order evolution from a high-entropy matrix: Understanding and predicting paths to low-temperature equilibrium. *J Am Ceram Soc.* **108**, e20223 (2025).
16. Tan, Y. *et al.* Phase-field study of precipitate morphology in epitaxial high-entropy oxide films. *Acta Materialia* **286**, 120721 (2025).
17. Sivak, J. T. *et al.* Discovering High-Entropy Oxides with a Machine-Learning Interatomic Potential. *Phys. Rev. Lett.* **134**, 216101 (2025).
18. Almishal, S. S. I. *et al.* Transparent Correlated Metallic Perovskites with Conducive Chemical Disorder.
19. Kotsonis, G. N. *et al.* Fluorite-structured high-entropy oxide sputtered thin films from bixbyite target. *Applied Physics Letters* **124**, 171901 (2024).
20. Kotsonis, G. N. *et al.* Property and cation valence engineering in entropy-stabilized oxide thin films. *Phys. Rev. Mater.* **4**, 100401 (2020).
21. Chae, S., Yoo, S., Kioupakis, E., Lu, W. D. & Heron, J. T. Perspective: Entropy-stabilized oxide memristors. *Applied Physics Letters* **125**, 070501 (2024).
22. Yoo, S. *et al.* Efficient data processing using tunable entropy-stabilized oxide memristors. *Nat Electron* **7**, 466–474 (2024).
23. Mazza, A. R. *et al.* Designing Magnetism in High Entropy Oxides. *Advanced Science* **9**, 2200391 (2022).
24. Mazza, A. R. *et al.* Embracing Disorder in Quantum Materials Design. *Applied Physics Letters* **124**, (2024).
25. Yan, J. *et al.* Orbital degree of freedom in high entropy oxides. *Phys. Rev. Materials* **8**, 024404 (2024).
26. Sarkar, A. *et al.* High entropy oxides for reversible energy storage. *Nature Communications* **9**, 3400 (2018).
27. Meisenheimer, P. B. & Heron, J. T. Oxides and the high entropy regime: A new mix for engineering physical properties. *MRS Advances* **5**, 3419–3436 (2020).
28. Schweidler, S. *et al.* High-entropy materials for energy and electronic applications. *Nat Rev Mater* **9**, 266–281 (2024).

# Supporting Information

**Note 1:** Figure S1 displays XRD patterns for bulk $Ca_x(MgCoNiCuZn)_{1-x}O$ with Ca concentrations from 4 to 16.6% (equimolar concentration). At Ca concentrations ≥5.5%, two peak sets are visible: one reminiscent of MgCoNiCuZnO, and one with CaO (111), (200), and (220) reflections indicating a CaO-rich phase. At 4% Ca, only one rock salt peak set is visible, indicating a single-phase material.

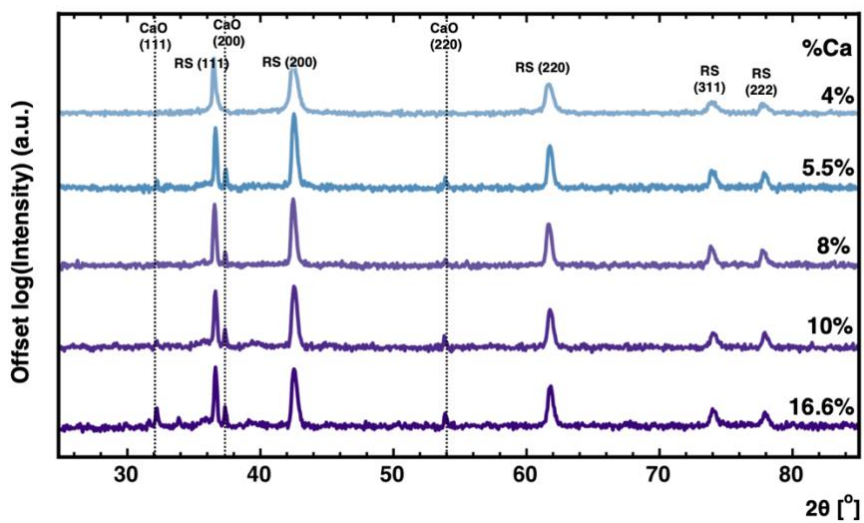

Figure S1: Bragg-Brentano High-Definition X-ray diffractometry (BBHD XRD) scans for bulk Cu-containing $Ca_x(MgCoNiCuZn)_{1-x}O$ with varying Ca concentration. Dotted lines are provided to clearly mark CaO peaks.

**Note 2:** Chemical compositions for each studied system are verified using X-ray fluorescence (XRF). To demonstrate calibrated XRF resolution, Figure S2 shows spectra for the endmember MgCoNiCuZnO and for $Ca_{0.05}(MgCoNiCuZn)_{0.95}O$. In the Ca-containing sample, distinct Ca $K_\alpha$ and $K_\beta$ peaks are clearly visible at approximately 3.69 keV and 4.01 keV, respectively. The intensity of these peaks then enables precise quantification of Ca content, confirming ~5 at % Ca incorporation in this case.

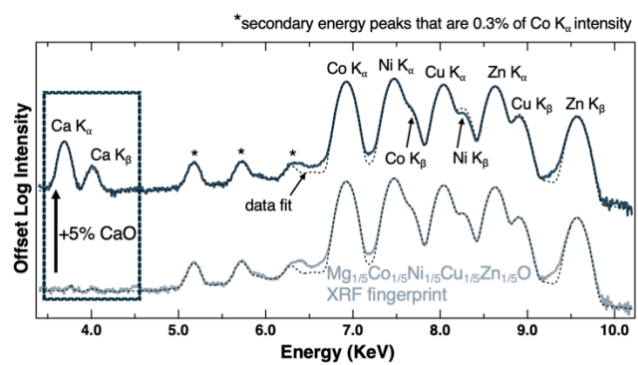

Figure S2: X-ray fluorescence spectra for MgCoNiCuZnO and $Ca_{0.05}(MgCoNiCuZn)_{0.95}O$.

**Note 3:** Figure S3 displays XRD patterns for bulk $Ca_x(MgCoNiZn)_{1-x}O$ with Ca concentrations 4-20%. At Ca concentrations ≥5%, two peak sets are visible: one reminiscent of MgCoNiZnO, and one with CaO (111), (200), and (220) reflections indicating a CaO-rich phase. Below 5% Ca, only one rock salt peak set is visible for each pattern, suggesting a single-phase material.

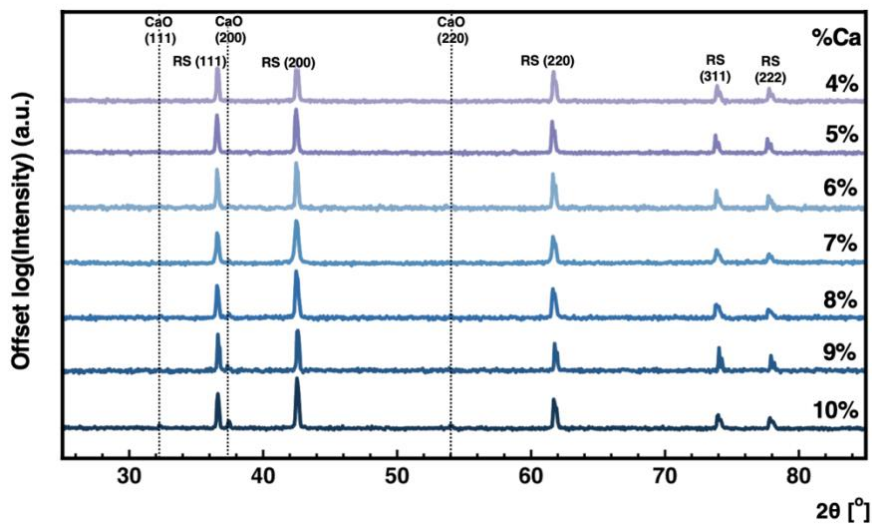

Figure S3: Bragg-Brentano High-Definition X-ray diffractometry (BBHD XRD) scans for bulk Cu-free $Ca_x(MgCoNiZn)_{1-x}O$ with varying Ca concentration. Dotted lines are provided to clearly mark CaO peaks.

**Note 4: Establishing favorable growth conditions**

To establish robust growth protocols, we systematically mapped the two key deposition variables: laser fluence and substrate temperature, across all Ca concentrations.

**Laser fluence.** Figures S4a (Cu-containing) and S4b (Cu-free) illustrate XRD scans collected at fluences from 0.9 to 2.5 J/cm² (increasing upward) at fixed 400°C substrate temperature. The film (002) and (004) reflections appear throughout the series; however: at fluence < 1.5 J/cm², lower kinetic energy yields additional low-intensity peaks, indicating mixed orientations, and at fluence > 2.0 J/cm² excess energetics again promotes secondary orientations. Therefore, we choose 1.87 J/cm², as the minimum fluence that consistently suppresses other orientations while also retaining high crystalline quality.

**Substrate temperature.** Figures S4c and S4d show analogous scans collected at temperatures from 200 to 600 °C at the optimal 1.87 J/cm² fluence. Single-orientation growth persists up to 500 °C. Above this threshold, a shoulder develops on the low-angle side of the (002) peaks, which could indicate a strain-relaxed Ca rich secondary phase, likely driven by enhanced cation diffusion at the high substrate temperature. We further hypothesize that the narrow temperature window for Ca-rich single-phase growth below 500 °C may be linked to a low-temperature oxidation of $Co^{2+}$ to $Co^{3+}$. The smaller ionic radius of $Co^{3+}$ could relieve lattice strain, freeing space for the larger

Ca²⁺ ions and thereby stabilizing the rock-salt structure. This mechanism warrants further investigation in future studies. Overall, we adopt 400°C as the best compromise between single phase and high crystallinity growth for both Cu-containing and Cu-free compositions, ensuring a direct comparison under identical conditions. Combined, these experiments define 1.87 J cm$^{-2}$ laser fluence and 400 °C substrate temperature as our favorable growth conditions across our composition space.

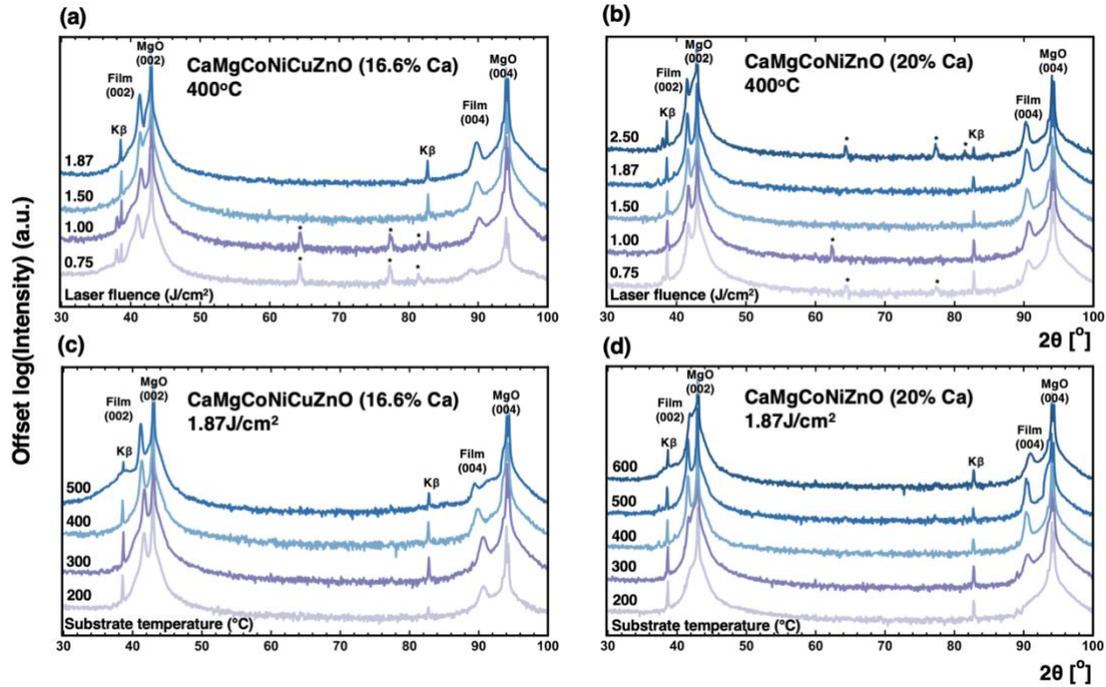

Figure S4: (a) Bragg-Brentano High-Definition X-ray diffractometry (BBHD XRD) scans for: (a) CaMgCoNiCuZnO with varying pulsed laser fluence, (b) CaMgCoNiZnO with varying pulsed laser fluence, (c) CaMgCoNiCuZnO with varying substrate temperature, and (d) CaMgCoNiZnO with varying substrate temperature. (*) denote a peak orientation other than the rock salt (002) or (004) peak.